\documentstyle[12pt,aaspp4]{article}

\def\etal{{\it et al.\thinspace}}

\def\fmag{\hbox{$.\!\!^m$}}

\begin{document}
\slugcomment{Submitted to {\it The Astronomical Journal}}

\lefthead{Carrasco et al.}
\righthead{Spectral Observations of Faint Markarian Galaxies II.}

\title{SPECTRAL OBSERVATIONS OF FAINT MARKARIAN GALAXIES
OF THE SECOND BYURAKAN SURVEY. II.}

\author{L. Carrasco\footnotemark
\footnotetext{Also Instituto de Astronomia, UNAM, Ensenada, M\'exico}
and H.M. Tovmassian}
\affil{Instituto Nacional de Astrof\'isica Optica y Electr\'onica,
Apartado Postal 51 y 216. C.P. 72000
Puebla, Pue., M\'exico \\
Electronic mail: carrasco@inaoep.mx, hrant@inaoep.mx}

\author{J.A. Stepanian}
\affil{Special Astrophysical Observatory RAS\ Nizhnij Arkhys,
Karachai-Cherkessia, 357147\\ Russia \\
Electronic mail: jstep@sao.ru}

\author{V.H. Chavushyan\footnotemark
\footnotetext{Also Special Astrophysical Observatory RAS, Russia}
and L.K. Erastova\footnotemark}
\footnotetext{Visiting Astronomer at the Special Astrophysical Observatory
RAS, Russia}
\affil{Byurakan Astrophysical Observatory, Byurakan, 378433 Armenia \\
Electronic mail: chavush@sao.ru and lke@sao.ru}

\author{J.R. Vald\'es}

\affil{Instituto Nacional de Astrof\'isica Optica y Electr\'onica,
Apartado Postal 51 y 216. C.P. 72000 \
Puebla, Pue., M\'exico \\
Electronic mail: jvaldes@inaoep.mx}

\begin{abstract}

We continue the program of spectroscopic observations of objects from the
Second Byurakan Survey (SBS). This survey contains more than 1300 galaxies
and 1700 star-like objects with $ m_{pg}<19\fmag5$. Our work is aimed
towards the construction of a complete sample of faint Markarian galaxies.
Here, we present spectroscopic data for 43 galaxies. Amongst them six new
Seyfert galaxies are found, namely two Sy 1 type (SBS~1343+544 and
SBS~1433+500), one Sy 2 type (SBS~1620+545) and three likely Sy 2 type
galaxies (SBS~1205+556, SBS~1344+527, SBS~1436+597). SBS~1343+544 falls
into the luminosity gap between low-redshift QSOs and high luminosity
Sefert galaxies. In the sample studied here, another 36 emission-line
galaxies were spectroscopically confirmed. Thus far, 102 SBS galaxies
brighter than 17\fmag5 have been observed with the Cananea 2.1m GHO
telescope.

The apparent magnitude and redshift distributions, the spectral
classification, the relative intensities of emission lines, and other
parameters, as well as slit spectra for all 43 observed galaxies are
presented.

\end{abstract}

\section{INTRODUCTION}

In order to built a complete sample of Markarian galaxies up to $
m_{pg}<17^{m}$ contained in the Second Byurakan Survey sample (Markarian
\& Stepanian \markcite{a4} 1983, \markcite{a5} 1984a,\markcite{a6} b;
Markarian, Stepanian \& Erastova \markcite{a7} 1985, \markcite{a8} 1986;
Stepanian, Lipovetsky \& Erastova \markcite{a10} 1988, \markcite{a11}
1990, Stepanian  \markcite{a13} 1994), a follow-up spectral study of the
selected objects is required. Our work is aimed toward this goal.

While high resolution spectral observations of faint objects are being
carried out, primarily with the 6 m telescope at SAO in Russia (Stepanian
\etal \markcite{a12} 1993 and references therein), observations of
relatively brighter objects are carried out with 2.1 m GHO telescope in
M\'exico. So far, hundreds of QSOs, Seyfert galaxies, Emission Line
Galaxies (ELGs), Blue Compact Dwarf Galaxies (BCDGs), as well as some
peculiar stars have already been found in the SBS sample.

This long term project shall allow us to compile the first complete sample
of faint ($ m_{pg}<17^{m}$) Seyfert and AGN galaxies. Such sample, shall
also allow us to construct the Luminosity Function of AGN and UVX galaxies
out to a distance of about 500 Mpc.

In a previous paper (Carrasco \etal 1997, \markcite{a2} Paper I) the
results of the follow-up spectroscopy  of 59 SBS galaxies were presented.
In that subsample, five new Seyfert galaxies were found and 51 emission
line galaxies were confirmed.

In the present paper we report on the results of the follow-up
spectroscopic observations of another 43 relatively bright galaxies
($15\fmag0-17\fmag5$) from the SBS sample. These observations make part of
an ongoing effort to obtain the data required by the scientific goals
outlined above. There is also a general interest in increasing the sample
of known Seyfert and AGN galaxies.

\section{OBSERVATIONS}

For our project, two observing runs in March and April 1997 were allocated
at Guillermo Haro Observatory of INAOE in Cananea, M\'exico. Observations
were carried out with the 2.1 m telescope  and the LFOSC spectrophotometer
equipped with a $600\times400$ pixel CCD. A set up covering the spectral
range of 4000-7100\AA\ was adopted. The effective instrumental spectral
resolution was about 11\AA.

The journal of  observations is presented in Table 1. In consecutive
columns the following data are listed: 1 -- the SBS designation (1950.0
epoch) 2-3 -- the 2000.0 epoch coordinates as measured by Stepanian \etal
\markcite{a14} (in preparation). These are accurate to about $\pm1$
arcsec, 4 -- an eye estimated $m_{pg}$  magnitude with an accuracy of
about $\pm0\fmag5$, 5 -- the SBS spectral class, 6 --the date of
observation, 7 -- the exposure time, 8 -- an alternative designation of
the object when available.

\section{DATA REDUCTION AND RESULTS}

The usual data reduction procedures -- cosmic ray hits removal, bias
and flat field corrections, wavelength linearization and flux calibration
-- were carried out with the IRAF reduction package for observations,
carried out in March, 1997. While observations carried out in April, 1997
were reduced with software packages developed at the Special Astrophysical Observatory, Russia (Vlasyuk \markcite{a16} 1993).

The integrated emission line fluxes were determined with the help of the
spectral analysis software package developed by Vlasyuk (private
communication) at the Special Astrophysical Observatory. This software
determines the best-fit Gaussian profile for every line and is capable of
deblending closely spaced lines, as is the case of ${H_\alpha}$~6563~\AA\
and the $[NII]$ ~6548, 6583~\AA\ lines. In our case, this blending is due
to the combined effects of the spectral resolution of the spectrometer and
the intrinsic width of the emission lines of the objects.

For every galaxy in our sample, the value of the dust reddening
coefficient $c(H_{\beta})$ was determined from the observed ratio of
$I({H_\alpha})/I(H_{\beta})$, assuming that the intrinsic ratio of
$F(H_{\alpha})/F(H_{\beta})$ is given by:

\begin{center}

$F(H_{\alpha})/F(H_{\beta})=I(H_{\alpha})/I(H_{\beta})\times
10^{c(H\beta)f(\lambda)}$

\end{center}

where ${f(\lambda)}$ is listed by Kaler \markcite{a3} (1976) for a
standard galactic reddening law (Whitford \markcite{a17} 1958).

The values of ${c}(H_{\beta})$ were computed by assuming that the
intrinsic ratio of $F(H_{\alpha})/F(H_{\beta})=2.85$ for narrow line
emission galaxies, and is equal to 3.1 for Sy 2's, and for narrow line
components of Sy 1.5 (Veilleux \& Osterbrock, \markcite{a15} 1987).

The results of spectral observations are presented in Table 2, in which
the following data are given:

1 -- The SBS designation.

2 -- The emission line redshift, derived as mean value of redshifts of
strong emission lines, corrected for solar motion:

\centerline{$\Delta z=0.001\sin l^{II}\cos b^{II}$}

3 -- The absolute magnitude $M_{pg}$ given by the expression:

\centerline{$M_{pg}=m_{pg}-5\log(z)-43.01-0.24~cosec~b^{II}$}

for~~$H_ {0}=75~km~s^{-1}~Mpc^{-1}$.

4 -- The reddening coefficient $ c(H_{\beta})$.

5 -- The $(H_{\beta})$ - equivalent width $EW(H_{\beta})$.

6-10 -- The logarithms of the observed and reddening corrected relative
intensities, normalized to $H_{\beta}$ or to $H_{\alpha}$ when $H_{\beta}$
is absent. The uncertainty in intensity ratios for strong emission lines
reported here is less than 30\%.

11 -- Spectral type. The Seyfert type, when marked ":" refers to an 
ambiguous spectral classification.

In the sample studied here, two Seyfert 1 type galaxies -- SBS~1343+544
and SBS~1433+500 -- were found. The spectrum of SBS~1343+544 shows broad
hydrogen emission lines, $H_{\beta}$ ($FWHM\sim8000$ km/s) and
$H_{\gamma}$ ($FWHM\sim7000$ km/s). While in the spectrum of SBS~1433+500
the broad emission line $FWHM$  $H_{\beta}\sim5000$ km/s is evident. In
both cases there is narrow [O III] emission.

SBS~1620+545 -- has a spectrum typical of Seyfert 2 type galaxies. Three
other objects, SBS~1205+556, SBS~1344+527 and SBS~1436+597, most likely
are also Seyfert 2 type galaxies.

Mkn~1488 (SBS~1359+521C), presents in its spectrum only the $MgIb$ and
$NaD$ absorption lines.

The magnitude and redshift distributions of the entire sample of SBS
galaxies and those of the subsample observed in Cananea (including data
from Paper I), are presented in Figures 1 and 2 respectively.

\placefigure{fig1}

\placefigure{fig2}

The diagnostic classification diagrams based on emission line ratios
(Veilleux and Osterbrock 1987) for the studied objects, including the
objects reported in Paper I, are presented in Figures 3(a) and 3(b).

\placefigure{fig3}

Plots of the spectra of the observed objects are presented in Figures 4
through 6.

\placefigure{fig4}
\placefigure{fig5}
\placefigure{fig6}

\section{CONCLUSIONS}

The results of the spectrophotometric observations of 43 SBS galaxies made
with the Cananea 2.1 m telescope in March 1997 and April 1997 are presented.

Among the objects observed, three new Seyfert galaxies were found. These
are two Seyfert 1 type galaxies, SBS~1343+544 and SBS~1433+500, one
Seyfert 2 type galaxy, SBS~1620+545. Another three galaxies --
SBS~1205+556, SBS~1344+527, and 1436+597 -- are probably also Seyfert 2
type objects. In total 42 emission-line galaxies, and one absorption line
galaxy were spectroscopically confirmed.

SBS~1343+544 falls in the gap between low-redshift QSOs and higher
luminosity Seyfert galaxies, probably meaning that it is a transition
object.

Spectral classification, redshifts, relative intensities of the prominent
emission lines, as well as other parameters were determined for all 42
emission line galaxies.

\section{ACKNOWLEGEMENTS}

We thank V. Vlasyuk and A. Burenkov for allowing us to use their data
analysis package. VHCh thanks INAOE's staff for their hospitality during
his visit. This work has been partially supported by CONACYT research
grants No. 211290-5-1430PE and 211290-5-0009PE. JAS, VHCh and LKE were
partially supported by the research grants No. 97-02-17168 and 1.2.2.2
from the Russian Foundation for Basic Research and from State Programm
"Astronomy" respectively. We also thank J. Miramon, G. Miramon, S. Noriega
and J. Genis for excellent assistance and technical support at the
telescope.

\begin{table}
\dummytable\label{tbl-1}
\end{table}

\begin{table}
\dummytable\label{tbl-2}
\end{table}

\newpage

\begin{center}
{\bf Figure Captions}
\end{center}

\figcaption [carrasco.fig1.ps] {The histograms of the stellar magnitude
distributions of the whole SBS sample and of the galaxies observed in
Cananea.\label{fig1}}

\figcaption [carrasco.fig2.ps] {The histograms of the redshift
distributions of the SBS sample with available spectroscopic data and of
the galaxies observed in Cananea.\label{fig2}}

\figcaption [carrasco.fig3.ps] {Emission line ratio classification
diagrams. The solid line marks the boundary between "HII-region-like"
galaxies and AGNs.\label{fig3}}

\figcaption [carrasco.fig4.ps] {Plots of the spectra of the observed
galaxies in relative flux units. Wavelengths in Angstroms.\label{fig4}}

\figcaption [carrasco.fig5.ps] {Plots of the spectra of the observed
galaxies in relative flux units. Wavelengths in Angstroms.\label{fig5}}

\figcaption [carrasco.fig6.ps] {Plots of the spectra of the observed
galaxies in relative flux units. Wavelengths in Angstroms.\label{fig6}}

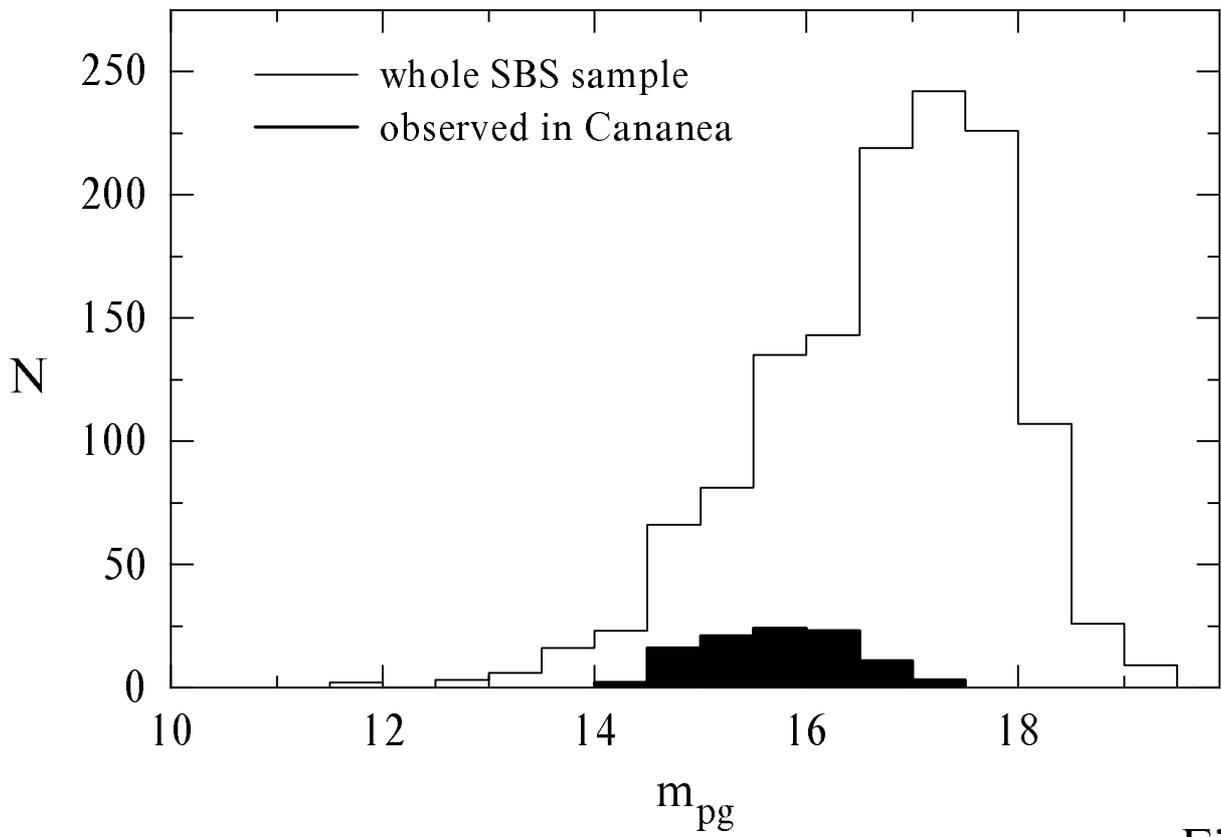
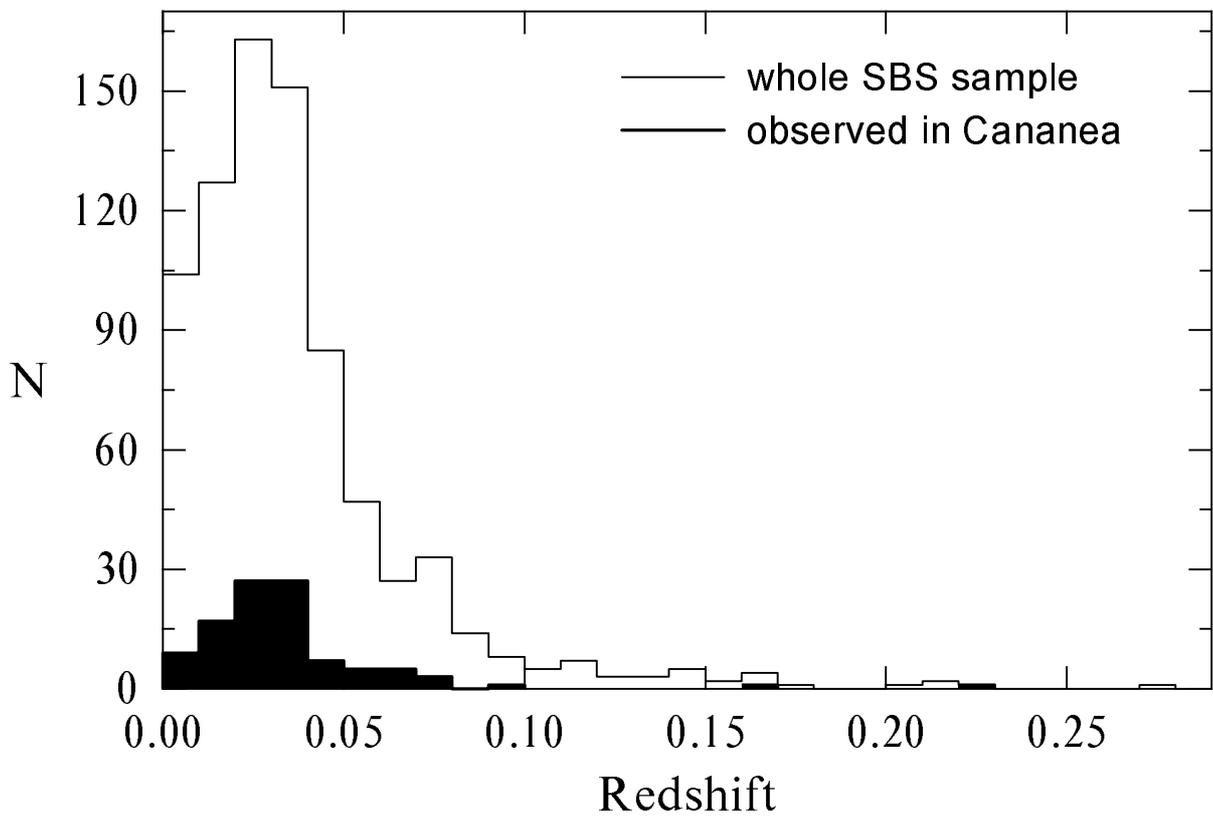
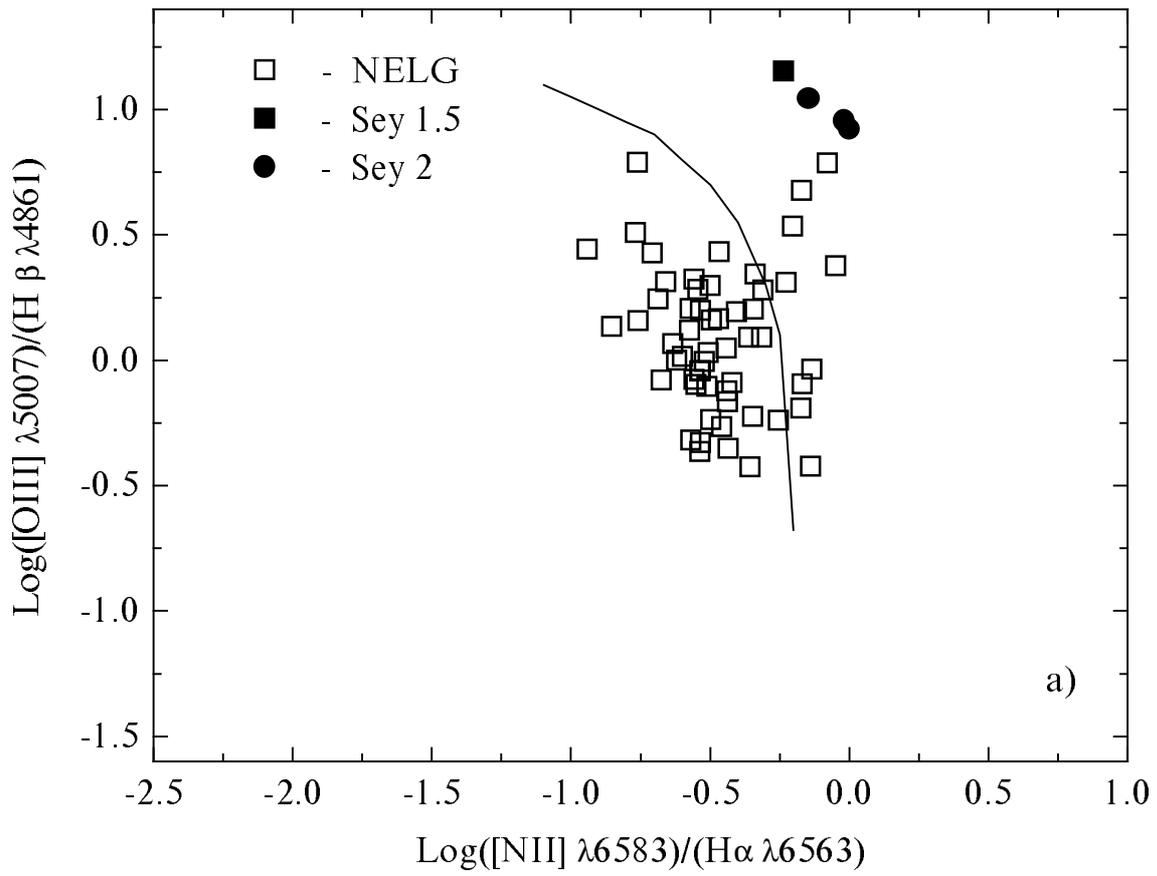
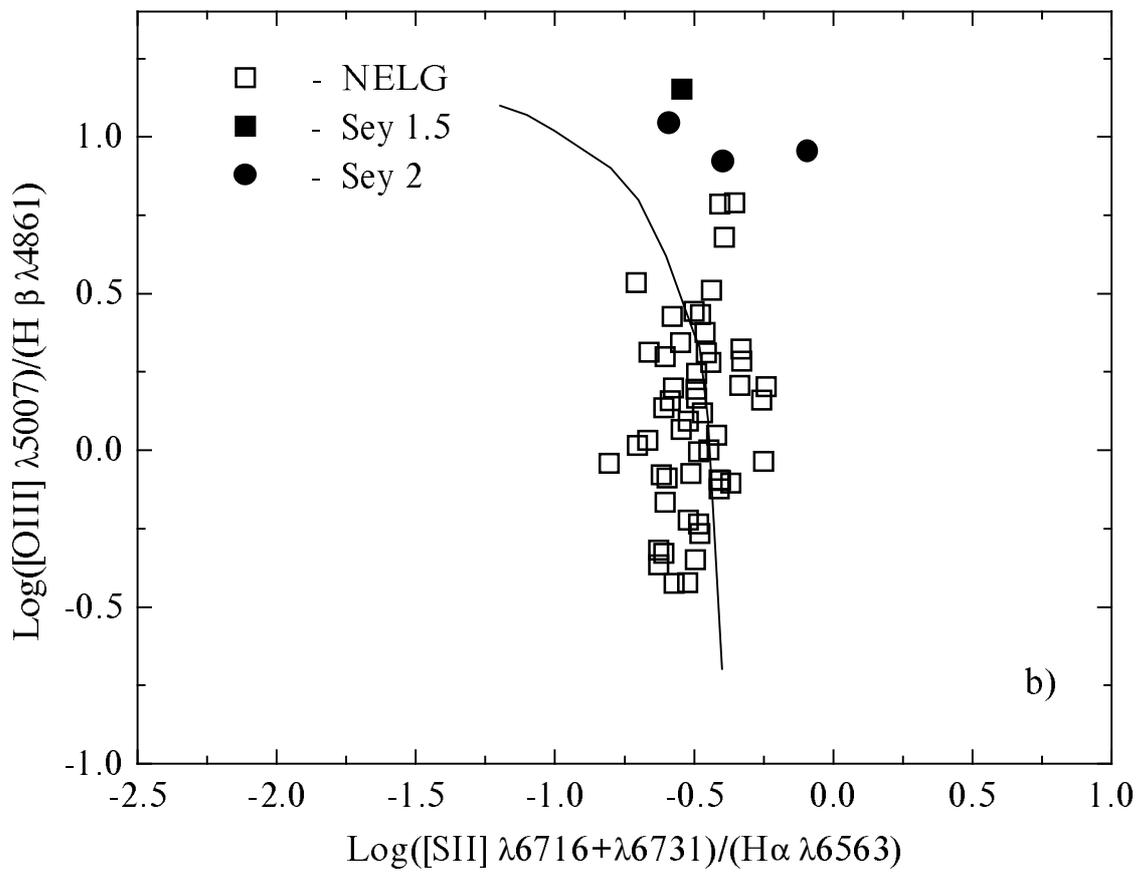
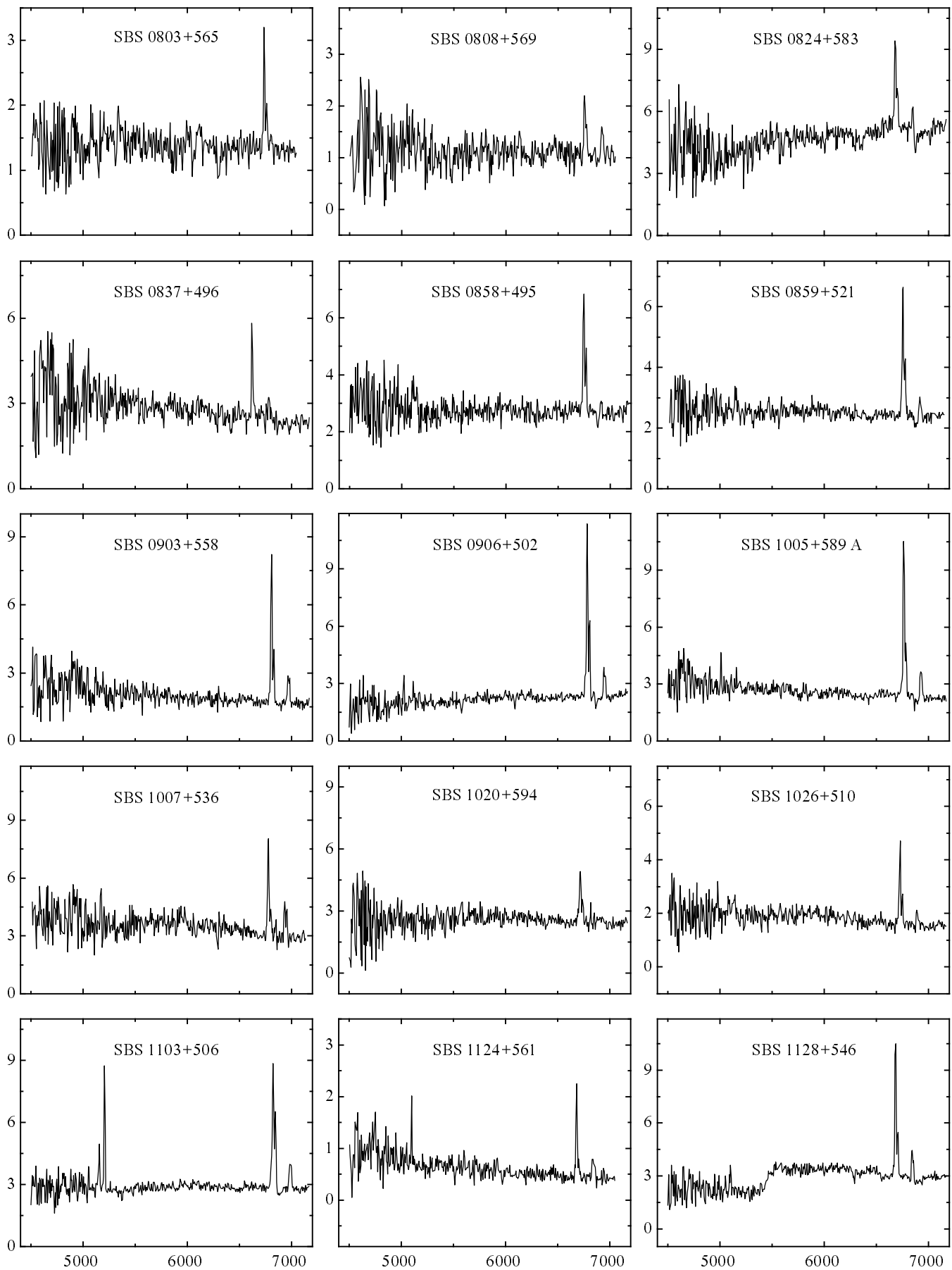
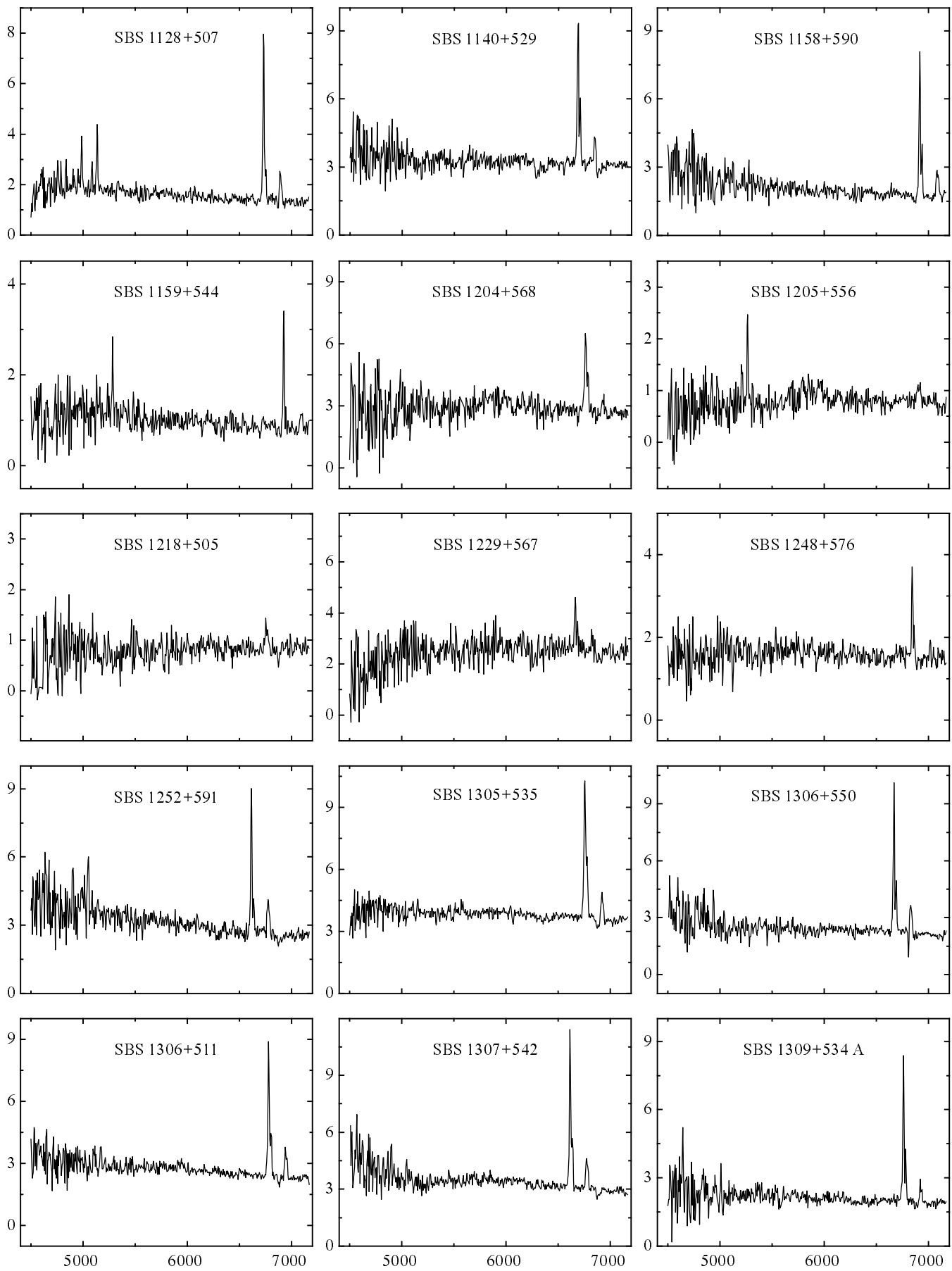
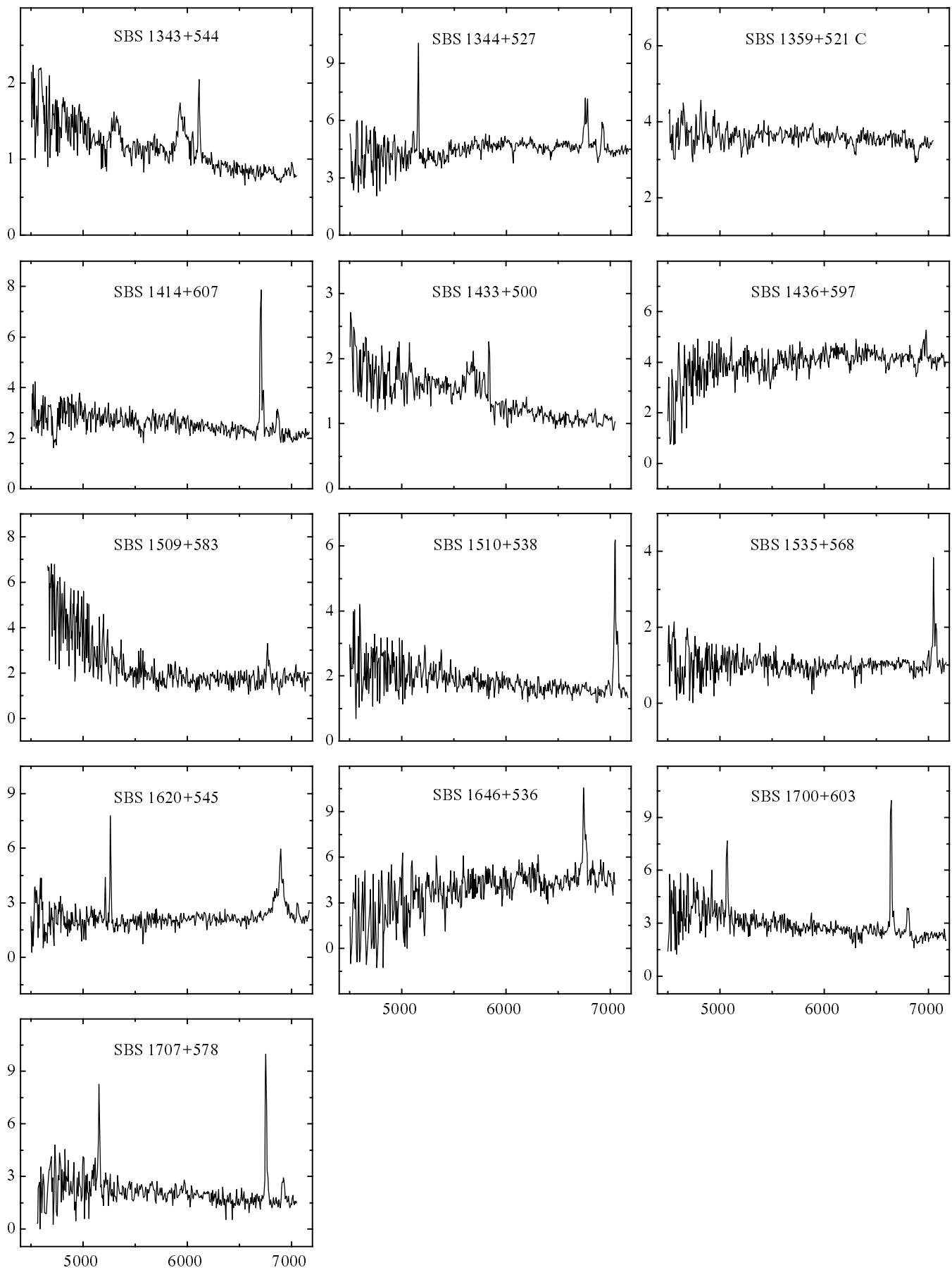
\end{document}